# Originality and the Future of Copyright in an Age of Generative AI

*Mark Fenwick and Paulius Jurcys*

**Abstract**

This paper takes the occasion of French DJ David Guetta's use of generative AI tools to create lyrics and a voice in the style of Eminem, which he then used in one of his concerts, as the basis for an exploration of the shifting meaning of creativity and originality in the age of generative AI. Our main contention is that the Guetta form of creativity with generative AI tools differs in certain important respects from what has come before. The paper describes an iterative, dynamic process of *conception, prompting, generation, refining,* and *deployment* to characterise creativity in this context. Nevertheless, we contend that copyright – specifically the concept of originality as articulated in US federal law – is a sufficiently durable legal mechanism that can manage these new cultural forms, and that the two basic requirements of modern copyright law (a tangible medium of expression and a modest degree of creativity) remain relevant in identifying the scope of legal protection.

   The paper argues that the David Guetta story reveals something more general about creativity in a digital age, namely that while hybrid-networked (i.e., human – corporate – machine) *creators* have always created hybrid-networked cultural *forms* (i.e., creations that blend human and technology-constituted elements), such hybridity becomes increasingly visible and complex in the context of a new world of generative AI. At the very least, earlier – and influential – models of creativity as human-driven involving creation *ex nihilo* become harder to sustain in a new age of generative AI. But this does not mean copyright or notions of originality are redundant or that copyright law cannot accommodate Guetta and other cases.

   Such an account seems important as it challenges the hegemonic and reductive view that AI "generates" artistic works autonomously and avoids reducing the copyright issues raised by such creative works to the related but distinct question of whether learning models rely on copyrighted data. As such, copyright law should remain an important mechanism to facilitate genuine creators who are using AI systems in innovative and unique ways to push the boundaries of their creativity.

**Keywords**

AI, generative AI, copyright, creativity, originality, artificial intelligence, data, machine learning, Feist, originality, David Guetta, ChatGPT, input, output

# Table of Contents

# 1. "Let Me Introduce You to … Emin-AI-Em"

In February 2023, one of the most prolific dance music producers in the world, French DJ, David Guetta, posted on his personal Instagram account a short clip from one of his latest concerts.[1] In the clip, you can hear what appears to be Eminem's voice saying, "This is the future rave sound; I'm getting off the underground," accompanied by energetic dance music. Future Rave, or FR, is a genre of electronic music associated with Guetta and Danish DJ, Morten Breum.

Guetta explained that this was something he "made as a joke," but it worked "so good that I could not believe it."[2] He described how he had discovered several generative AI websites that could be used to "create" lyrics and voices. One of them helped Guetta generate lyrics in the style of any artist, so he typed the prompt, "Write a verse in the style of Eminem about the Future Rave." He then used another generative AI website that recreated Eminem's voice. Finally, he combined the words in Eminem's voice with some music. The Instagram clip ends with Guetta talking about the effect of his "creation" on the audience: "People went nuts."

Several commentators observed that this was not the first time Guetta paid tribute to Eminem – in 2019, he played Eminem's "Lose Yourself" at the MDL Beast Festival.[3] But on that occasion, and in contrast to his later usage, he played Eminem's original music (and voice).[4] However, soon after making the initial Instagram post, Guetta gave an interview where he explained that he would not release the mix publicly because of his respect for Eminem and the ambiguous legal framework concerning copyright entitlements.[5] In another interview to BBC, he stated that, "I'm sure the future of music is in AI. For sure. There's no doubt. But as a tool. … Nothing is going to replace taste. What defines an artist is, you have a certain taste, you have a certain type of emotion you want to express, and you're going to use all the modern instruments to do that."[6]

---

[1] See <https://www.instagram.com/p/CoNqQuFqIHZ/> accessed 12 February 2023.

[2] Ibid.

[3] See, eg, 'David Guetta plays Eminem's "Lose Yourself"' (*Southpawers*, 5 February 2023) <https://southpawers.com/2023/02/05/david-guetta-eminem-ai/> accessed 12 February 2023.

[4] Southpawers, 'MDL Beast Fest 2019' <https://www.youtube.com/watch?v=BTA6LmwhNYk&t=2s> accessed 12 February 2023.

[5] Sam Roche, 'David Guetta: "If you have terrible taste, your music is still gonna be terrible, even with AI"' (*MusicTech*, 24 July 2023) <https://musictech.com/news/david-guetta-on-ai/> accessed 30 August 2023.

The emergence and proliferation of generative AI technologies in the first half of 2023 spurred a huge global debate among lawyers, policymakers, and technologists, as well as the general public.[7] The rapid development of generative AI even caused waves of public outcry and calls to halt the development of AI for several months or at least until some legal, technological, and ethical guardrails are established.[8] However, since many generative AI tools are open source, technology entrepreneurs rushed to embrace the new wave of innovation and start building specific solutions for diverse use cases across enterprise and consumer settings.[9]

---

[6] 'David Guetta says the future of music is in AI' (*BBC*, 13 February 2023) <https://www.bbc.com/news/entertainment-arts-64624525> accessed 30 August 2023.

[7] See eg, Franklin Graves, 'Accelerated Innovation: In Less Than a Year, We've Seen a Decade's Worth of AI and IP Developments' (*IP Watchdog*, 13 August 2023) <https://ipwatchdog.com/2023/08/13/accelerated-innovation-less-year-weve-seen-decades-worth-ai-ip-developments/id=164842/> accessed 29 August 2023; 'Secretary-General Urges Security Council to Ensure Transparency, Accountability, Oversight, in First Debate on Artificial Intelligence' (United Nations, 18 July 2023) <https://press.un.org/en/2023/sgsm21880.doc.htm> accessed 30 August 2023; Gideon Lichfield & Lauren Goode, 'The World Isn't Ready for the Next Decade of AI' (*Wired*, 16 August 2023) <https://www.wired.com/story/have-a-nice-future-podcast-18/> accessed 30 August 2023; Colleen Walsh, 'How to think about AI: Delving into the legal and ethical challenges of a game-changing technology' (*Harvard Law Bulletin*, Summer 2023) <https://hls.harvard.edu/today/how-to-think-about-ai/> accessed 29 August 2023.

[8] The Future of Life Institute, 'Pause Giant AI Experiments: An Open Letter' <https://futureoflife.org/open-letter/pause-giant-ai-experiments/> accessed 29 August 2023; Mustafa Suleyman a& Michael Bhaskar, *The Coming Wave* (Crown 2023); Ben Tarnoff, 'Weizenbaum's nightmares: how the inventor of the first chatbot turned against AI' (*The Guardian*, 25 July 2023) <https://www.theguardian.com/technology/2023/jul/25/joseph-weizenbaum-inventor-eliza-chatbot-turned-against-artificial-intelligence-ai?CMP=Share_iOSApp_Other> accessed 29 August 2023.

[9] Sam Altman: 'The reason people love this stuff [AI] is because it's providing real utility. And that doesn't come along too often.' See 'A conversation with OpenAI CEO Sam Altman' (YouTube, 18 May 2023) <https://www.youtube.com/clip/UgkxNvSNAau93YdpBGWEe9rUGyhajFfi9jsb> accessed 30 August 2023. It is expected that by the end of 2023, there will be appr. 12,000 startups working on AI projects. see For an updated list of such AI solutions, see: <https://theresanaiforthat.com/> accessed 30 August 2023.

From a copyright law and regulatory perspective, generative AI technologies raise important questions around four main areas. The first area of uncertainty relates to the legality of data scraping and using publicly available information (including content that is both protected by copyright laws and content in the public domain) to train machine learning and AI models.[10] Here, the question arises whether the scraping of data done by machines should be treated differently from the retrieval of the same information by humans.[11] Should permissions to use or licenses be obtained by entities that are scraping data for machine learning purposes?[12] Should the use of data scraped from the Internet be deemed to be a copyright infringement and thus not permissible?[13] In common law jurisdictions, for instance, one of the main issues is whether data scraping for machine learning models could be justified under the copyright law doctrine of fair use.[14] Then there are questions about remuneration.[15]

---

David G Widder, Sarah West & Meredith Whittaker, 'Open (For Business): Big Tech, Concentrated Power, and the Political Economy of Open AI'(2023) <https://papers.ssrn.com/sol3/papers.cfm?abstract_id=4543807> accessed 30 August 2023.

[10] 'AI is setting off a great scramble for data' (*Economist*, 13 August 2023) <https://www.economist.com/business/2023/08/13/ai-is-setting-off-a-great-scramble-for-data> accessed 29 August 2023.

[11] Mark A Lemley & Bryan Case, 'Fair Learning' (2021) 99 Tex L Rev 743

[12] Sharon Goldman, 'Generative AI datasets could face a reckoning' (*Venture Beat*, 21 August 2023) <https://venturebeat.com/ai/generative-ai-datasets-could-face-a-reckoning-the-ai-beat/> accessed 29 August 2023.

[13] See Kali Hays, 'OpenAI now tries to hide that ChatGPT was trained on copyrighted books, including J.K. Rowling's Harry Potter series' (*Business Insider*, 15 August 2023) <https://www.businessinsider.com/openais-latest-chatgpt-version-hides-training-on-copyrighted-material-2023-8> 29 August 2023; Winston Cho, 'Scraping or Stealing? A Legal Reckoning Over AI Looms' (*Hollywood Reporter*, 22 August 2023) <https://www-hollywoodreporter-com.cdn.ampproject.org/c/s/www.hollywoodreporter.com/business/business-news/ai-scraping-stealing-copyright-law-1235571501/amp/> accessed 29 August 2023.

[14] Lemley & Case, supra n 11, 748; Peter Henderson, Xuechen Li, Dan Jurafsky, Tatsunori Hashimoto, Mark A. Lemley, Percy Liang, 'Foundation Models and Fair Use' (2023) <https://arxiv.org/abs/2303.15715> accessed 30 August 2023.

[15] Martin Senftleben, 'Generative AI and Author Remuneration' <https://papers.ssrn.com/sol3/papers.cfm?abstract_id=4478370> accessed 30 August 2023; Miranda

The second major area of controversy surrounding generative AI technologies relates to data privacy and image rights issues. What if the data scraped from the Internet contains some personal information of individuals?[16] Could such data be used to train machine learning models? And which laws should be applied to such activities? The law of the state where the allegedly infringing AI company is based or is operating? Or the law of the place where individuals' personal information is being used? Or some other country's law? Can anyone create chatbots and avatars representing celebrities' and influencers' personalities and use the likeness to offer certain commercial products (e.g., a virtual representation of a notorious figure who acts as my shopping assistant)?[17] Furthermore, there seems to be no legal framework that would govern the interaction between *multiple* AI systems and agents with one another or third parties.[18]

The third area of legal uncertainty pertains to the large language models themselves. Different stakeholders have called for greater transparency in understanding what data is used to train

---

Nazzaro, 'Diller confirms plans for legal action over AI publishing' (*The Hill*, 16 JUly 2023) <https://thehill.com/policy/technology/4100533-diller-confirms-plans-for-legal-action-over-ai-publishing/> accessed 30 August 2023.

[16] Will Oremus, 'Meet the hackers who are trying to make AI go rogue' (*Washington Post*, 8 August 2023) <https://www.washingtonpost.com/technology/2023/08/08/ai-red-team-defcon/> accessed 30 August 2023; ICO, 'Joint statement on data scraping and data protection' <https://ico.org.uk/about-the-ico/media-centre/news-and-blogs/2023/08/joint-statement-on-data-scraping-and-data-protection/> accessed 30 August 2023.

[17] See <https://goshopwith.ai/chat> accessed 30 August 2023.

[18] Joon Sung Park, et al, 'Generative Agents: Interactive Simulacra of Human Behavior' (2023) <https://arxiv.org/abs/2304.03442> accessed 30 August 2023; Quingyun Yu, et al., 'AutoGen: Enabling Next-Gen LLM Applications via Multi-Agent Conversation Framework' <https://arxiv.org/pdf/2308.08155.pdf> accessed 30 August 2023; Aneesh Tickoo, 'Google DeepMind and the University of Tokyo Researchers Introduce WebAgent: An LLM-Driven Agent that can Complete the Tasks on Real Websites Following Natural Language Instructions' (*Markettechpost*, 29 July 2023) <https://rb.gy/qs5cv> accessed 30 August 2023; in August 2023, Andreesen Horowitz launched "AI-town", a JS starter kit for customizing your own "AI simulation" where AI characters live, chat and socialize: <https://github.com/a16z-infra/AI-town> accessed 30 August 2023.

large language models, and the core technology takes on the character of a "black box."[19] Much work from a technological, legal, and ethical perspective needs to be done to get a better insight into how LLMs work and what oversight mechanisms need to be in place.[20]

Finally, greater legal certainty is desirable with regard to the legality of outputs generated by individuals utilizing generative AI tools. Do such works created with generative AI tools meet the originality condition for the protection of works under copyrightlaw?[21] Who has what rights to outputs generated with generative AI tools?[22] In practice, having (any) rights to the output generated with the help of AI tools is a huge problem that will determine whether such works can be commercialized. At the time of writing, one of the main reasons for delays in the adoption of generative AI technologies by Fortune 1000 companies is uncertainty whether the outputs of machine learning models are infringing or not.[23] Then, there are questions about the relationship between the training data and the outputs.[24] Should the attribution of the rights to the output generated by an artist depend on what data was used to train the underlying model? Could output results be considered to be "derivative works" (from a copyright law perspective) of the training data?[25] If so, then there may be massive restrictions on how those outputs might be used.

---

[19] Carol Mullin Hayes, 'Generative Artificial Intelligence and Copyright: Both Sides of the Black Box' <https://papers.ssrn.com/sol3/papers.cfm?abstract_id=4517799> accessed 30 August 2023.

[20] 'Biden-Harris Administration Announces New Actions to Promote Responsible AI Innovation that Protects Americans' Rights and Safety' <https://www.whitehouse.gov/briefing-room/statements-releases/2023/05/04/fact-sheet-biden-harris-administration-announces-new-actions-to-promote-responsible-ai-innovation-that-protects-americans-rights-and-safety/> accessed 30 August 2023.

[21] Vincenzo Iaia, 'To Be, or Not to Be … Original Under Copyright Law, That Is (One of) the Main Questions Concerning AI-Produced Works' (2022) 71 GRUR Int 793.

[22] See e.g., Dan L Burk, 'Thirty-Six Views of Copyright Authorship, by Jackson Pollock' (2020) 58 Hous L Rev 263, 266; Ryan Benjamin & Elizabeth Rothman, 'Disrupting Creativity: Copyright Law in the Age of Generative Artificial Intelligence' Fla L Rev (forthcoming) <https://papers.ssrn.com/sol3/papers.cfm?abstract_id=4185327> accessed 29 August 2023; See e.g., David Newhoff, 'AI Machine Learning: Remedies Other Than Copyright Law?' <https://illusionofmore.com/ai-machine-learning-remedies-other-than-copyright-law/> accessed 30 August 2023.

[23] Paulius Jurcys, 'Event Recap: Augmenting Consumer Experiences in the Age of Data & AI' <https://www.prifina.com/blog/event-recap-augmenting-consumer-experiences-in-the-age-of-data-ai> accessed 29 August 2023.

In this paper, we suggest that the Guetta example constitutes a typical creative use case of generative AI in a contemporary context. Such generative AI increasingly finds applications across every domain of the creative industries.[26] From composing music, drafting text, writing software code and other technical documents rendering a room with a particular design or filled with a specific style of furniture[27] – generative AI permeates every area of creative work and defines digital culture, more generally.

---

[24] See eg, Matt Growcoot, 'Stability AI Boss Admits to Using 'Billions' of Images Without Consent' (*Petapixel*, 13 July 2023)
<https://petapixel.com/2023/07/13/stability-ai-boss-admits-to-using-billions-of-images-without-consent/> accessed 29 August 2023.

[25] Daniel J Gervais, 'AI Derivatives: the Application to the Derivative Work Right to Literary and Artistic Productions of AI Machines' (2002) 53 Seton Hall L Rev 1.

[26] For a general overview see Sonya Huang, Pat Grady and GPT-3, 'Generative AI: A Creative New World' <https://www.sequoiacap.com/article/generative-ai-a-creative-new-world/> (*Sequoia*, 19 September 2022) accessed 12 February 2023; for the investment landscape, see 'The state of generative AI in 7 charts' (*CB Insights*, 25 January 2023)
<https://www.cbinsights.com/research/generative-ai-funding-top-startups-investors/> accessed 12 February 2023.

[27] See eg, <https://interiorai.com/> accessed 12 February 2023; Mark WIlson, 'OpenAI's first acquisition is an AI design company' (*Fast Company*, 17 August 2023)
<https://www.fastcompany.com/90940634/openais-first-acquisition-is-an-ai-design-company> accessed 29 August 2023.

Judging by the initial reaction to recent developments in generative AI, the initial response seemed to be, at best, one of anxiety and, at worst, outright horror.[28] There are widespread concerns that these trends will result in machines taking over many aspects of work or worries about the capacity of the existing regulatory framework to deal with these unprecedented creative forms.[29] These developments certainly raise some interesting and important legal and philosophical questions about the meaning of creativity and the value of works created by individuals using generative AI tools. Are such works genuinely creative and original? What – if anything – is their value to society? How can we determine if such creations should be protected as creative works under copyright law? If so, how much protection should be afforded to such works created using generative AI tools? And, more dramatically, does this mark a new phase in human history, and will AI tools soon replace (or, in the doomsday account, eliminate) humans?[30]

Our intention here is to clarify one point about the degree and character of human involvement and the legal implications of that involvement. Firstly, and most obviously, it is essential to acknowledge that there is no such thing as a purely AI-generated work,[31] in the sense that some degree of prompting, refinement, and approval is required. Equally, some works with minimal human prompting are not of high value; therefore, copyright should be thin.[32] However, this recognition of the human/author in the loop has important implications. Most obviously, it chimes with the enterprise discourse around such products and systems, which frequently uses terms like co-piloting, co-creation, or companions to capture this dynamic element of prompting, editing, remixing, and approval in creating a final output.

---

[28] 'Expert AI as a Healthcare Superpower' (*A16z Podcast*, 10 January 2023) <https://rb.gy/lcayeo> accessed 12 February 2023; Gerard Baker, 'Is There Anything ChatGPT's AI 'Kant' Do?' (*The Wall Street Journal*, 13 February 2022) <https://rb.gy/hlbqpx> accessed 12 February 2023; Hugh Stephens, 'AI and Computer-Generated Art: Its Impact on Artists and Copyright' (*Hugh Stephens Blog*, 25 October 2022), <https://rb.gy/srndpb> accessed 12 February 2023; Stephen L Carter, 'Can ChatGPT Write a Better Novel Than I Can? (*Bloomberg*, 11 February 2023) <https://rb.gy/pnbrvx> accessed 13 February 2023.

[29] See e.g., Andy Kessler, 'AI's Growing Legal Troubles' (*The Wall Street Journal*, 30 July 2023) <https://www.wsj.com/articles/ais-growing-legal-troubles-section-230-publisher-class-action-9efaf374> accessed 29 August 2023.

[30] See eg, Allison Whitten, 'Me, Myself, and AI' (*Stanford Magazine*, July 2023) <https://stanfordmag.org/contents/me-myself-and-ai?sf179987774=1> accessed 29 August 2023; Shana Lynch, 'Will Generative AI Make You More Productive at Work? Yes, But Only If You're Not Already Great at Your Job' <https://stanfordmag.org/contents/me-myself-and-ai?sf179987774=1> accessed 29 August 2023.

[31] See eg, Emmanuel Salami, 'AI-Generated Works and Copyright Law: Towards a Union of Strange Bedfellows' (2020) 16 J Intell Prop L & Prac 124.

In thinking about the legal implications of this recognition of the human/author in the loop, we focus on one aspect of this issue, namely the meaning of "originality" in a copyright law context and the capacity of this concept as manifested in US federal law to accommodate the Guetta case and generative AI. Originality is a concept frequently referred to when talking about copyright, as it provides an important means to differentiate works considered socially valuable and worthy of copyright protection from those that lack this quality of originality. To be clear, it is not our intention to defend the status quo but merely to clarify the scope of copyright protections under contemporary conditions, focusing on US federal copyright law.

We limit our current analysis to US federal copyright law for three reasons: First, the United States (and the State of California in particular) is the place where most of the leading generative AI companies are headquartered.[33] Second, the United States is the main source of venture capital investments pouring into companies building AI technologies.[34] Finally, the fact that all of the major copyright lawsuits concerning generative AI are filed before US courts warrants the focus of this paper on the US copyright framework.[35]

---

[32] Here, it is important to note the concept coined by Rochelle C Dreyfuss, "if value then right" in her seminal article 'Expressive Genericity: Trademarks as Language in the Pepsi Generation' (1990) 65 Notre Dame L Rev 397, 405; for the applicability of this idea to works created with generative AI tools, see Lemley, supra n 61, 12-13.

[33] 42 out of 50 top AI companies are located in the US (35 in California), Ryan Heath, 'AI boom's big winners are all in four states' <https://www.axios.com/2023/07/24/ai-goldrush-concentrated-4-states> accessed 29 August 2023.

[34] Kyle Wiggers, 'VCs continue to pour dollars into generative AI' (*TechCrunch*, 28 March 2023) <https://techcrunch.com/2023/03/28/generative-ai-venture-capital/> accessed 29 August 2023; Anna Cooban, 'AI investment is booming. How much is hype?' (*CNN Business*, 23 July 2023) <https://www.cnn.com/2023/07/23/business/ai-vc-investment-dot-com-bubble/index.html> accessed 29 August 2023.

In this paper, we focus on the originality requirement; other issues – e.g., the legality of using publicly available data to train large language models, whether and how fair use might factor into this discussion[36] – are only briefly addressed in Section 5, which deals with broader issues that surround the debate on the issue of originality. Our reason for narrowing the scope and focusing on originality is motivated by our observation of several misconceptions about the degree of human involvement in the creative process where AI tools are used and how this issue is approached in the increasingly heated public debates.

For this reason, we begin our analysis in Section 2 by characterising the creative process when artists like David Guetta rely on generative AI tools in their creative processes. We offer a comparative illustration by contrasting these AI tools with a more traditional form of creation, such as painting a picture or writing a novel. This will highlight some of the distinctive features of creation in an age of generative AI. The question we then pose is whether the creative process with AI tools is any different from creativity where "traditional" creative tools are involved and whether this form of creation where generative AI tools are used represents something new. Our answer is that it does, but we need to be careful and precise in our characterisation of exactly what constitutes "newness" in the context of generative AI.

Section 3 turns to a brief discussion of extant copyright law in the United States. Two foundational distinctions – first, between ideas (which should remain free and not protected by copyright) and expressions (which can be subject to copyright protection) and second, the standard legal requirements of originality as a tool to differentiate works deserving of copyright protection – are introduced. The concept of originality will help us better address the "copyrightability" of works created using generative AI tools. This issue is addressed in Section 4.

---

[35] See eg, *PM v. OpenAI LP*, ND Cal, No 3:23-cv-03199 (complaint filed 28 June 2023); Tremblay v. OpenAI Inc., ND Cal, No 3:23-cv-03223 (complaint filed 28 June 2023); for up-to-date status of the Github Copilot litigation, see <https://githubcopilotlitigation.com/> accessed 30 August 2023; *Getty Images, Inc v Stability AI*, DC Del., No. 1:23-cv-00135 (complaint filed 3 February 2023)

[36] See Lemley & Casey, supra n 11, 743.

Section 5 explores how the human-in-the-loop concept connects with the broader discussion about the copyrightability of works created by human authors using generative AI tools and whether the assessment of originality of such works should include the lawfulness of the use of data for the training of the underlying large language models. Finally, Section 6 concludes.

Finally, we recognise that generative AI raises several other questions. For example, whether it makes a difference where works are created by humans or AI, and what are the interests of other participants in the ecosystem (e.g., an author who is upset that other people use generative AI tools the create "in the style of" the author), the justification of copyright, or the use of copyrighted material in machine learning models. These questions were not the subject of this paper and should be explored separately, as they are important.

## 2. A New Model of Creativity, Or Old Wine in New Bottles?

In the following, we aim to demystify what happens during the creative process, and we suggest that a "black box" approach that minimizes human involvement in AI-generated work is inadequate. To illustrate our point, we will rely on the example of David Guetta introduced at the beginning of this article. We first describe the process that characterises Guetta's usage of generative AI. Our suggestion is that his creative process is typical and a representative use case of generative AI. In elaborating this account, a contrast is made with a more typical case of creation, namely, an artist painting a picture. This comparison will allow a more precise clarification of the distinctiveness of creation in a world of generative AI and – crucially – a better understanding of the role of technology and a broader network of actors – legal, as well as natural persons – in all creation, but especially in a contemporary context.

*Figure 1. The Creative Process in an Age of Generative AI*

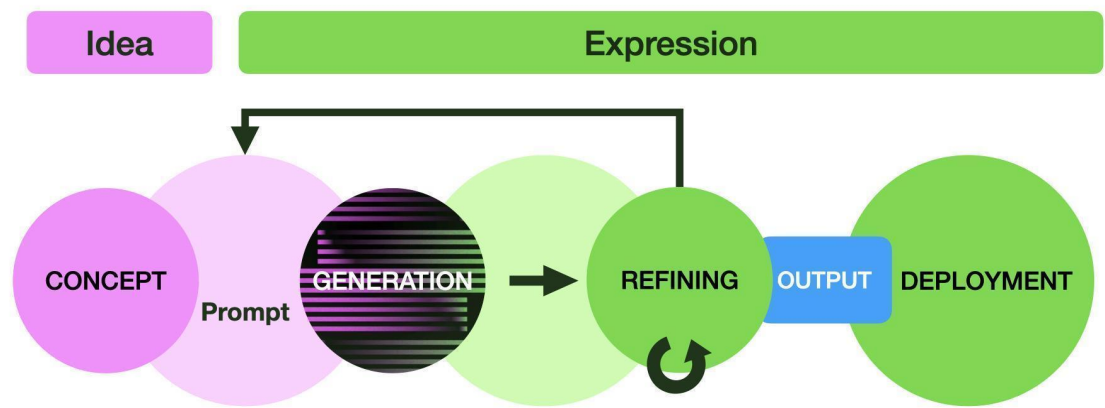

Five steps of David Guetta's usage of technology should be noted (see *Figure 1*).[37] Based on some *concept* of what he wanted to achieve (an iteration of FR music featuring the lyrics in the style of and the voice of Eminem); Guetta *prompted* the AI with a request to create a few lines of lyrics in the style of Eminem. Then, the AI – a piece of software utilising machine learning produced by a third party or parties – *generated* the requested text material and delivered the results back to Guetta. Another prompt was similarly used with a different AI system to generate the voice. Guetta *refined* the version of the piece by re-prompting, re-generating, editing, compiling, and integrating the results generated by the two different "AIs" with music creating an output, which – finally – was *deployed* in his DJ set.

A premise of the discussion is that this process of *conception*, *prompting*, *generation*, *refining*, and *deployment* characterises the creative use of generative AI. There are several aspects of this creative process that we would emphasise:

- ***Conception.*** Guetta started with an idea and knew what he wanted – Eminem-style lyrics and voices accompanying energetic four-to-the-floor FR – although he did not know, specifically, what he would get from the AI. He had a vision when he started, but he was not working with or from a fixed plan or script – i.e., he didn't know, precisely, where he was going. This is nothing new – all artists, our painter for example, presumably start the creative process with a similar vision or conception, however, minimal of what it is they are trying to achieve in a particular instantiation of the creative process, even if they don't always have a fixed conception of what it is they want to create.[38] In a creative context, intentions are important but malleable and incomplete and contingent upon context, especially previous works by dominant figures in the genre and tradition, as well as related genres and traditions.

---

[37] Other scholars have also tried to visualise the human-machine interactions in the creative process, see Jane C Ginsburg & Luke A Budiardjo, 'Authors and Machines' (2019) 34 Berkeley Tech L J 343 and P Bernt Hugenholtz and João Pedro Quintais, 'Copyright and Artificial Creation: Does EU Copyright Law Protect AI-Assisted Output?' (2021) 52 IIC 1190.

- *Prompting.* Crucially, generative AI is not – for the moment, at least – acting independently, in the sense that it requires external instruction, i.e., an input of some kind[39] (in this case, "Write a verse in the style of Eminem about the Future Rave)." This instruction was a self-conscious and deliberative choice on the part of a human creator aiming to materialise their concept. This second step does appear to be novel – there is nothing equivalent or comparable in the artist's case, even if certain aspects or conditions of prompting – engaging in some sort of background or preliminary research, for example – may be part of more traditional modes of creativity.

- *Generation*. The act of generation, in itself, is obviously not new – the painter paints, after all – and generating something would seem to be a necessary condition of all creative activity. Moreover – and this may be a slightly more contentious observation – the reliance on technology is not new either. A painter depends on simpler technology – brushes, paints, and paper – but *some* technology and by extension the producers of that technology are implicated in all creative processes and content generation. In an important sense, therefore, all creativity is a co-creation of human and machine and implicates the involvement of third parties. As such, technologies can be thought of as actors in the creative process, not in the sense that they take action, but in that they facilitate action and put action in motion and "make a difference in another agent's action."[40] In the context of his work on scientific invention, the French sociologist, Bruno Latour, talked of the need for a "flat ontology" in which we don't artificially over-emphasise the role of human beings in human activities (such "purification" is what he thought of as the arrogance of modernity) but rather see creativity as an *effect* of "networks" of human and machines. However, what is certainly distinctive in the Guetta case is the (nearly) complete delegation of the act of generation to a third party or parties (the generative AI software). The content that generative AI tools create depends on the prompts provided by the human author: the human author provides instructions and tasks for the generative AI assistant to deliver an iteration of the ideas that the human author is prompted to explore. And even though what the AI created or gave back to Guetta in its unedited form was not simply a copy of anything that already existed but something genuinely new and unique and that didn't exist before, such generated output is molded and shaped by initial prompts provided by a human author. The generated content is in some obvious sense new, even if the act of creating the lyrics and the voice, the AI software relied on enormous quantities of data accumulated, typically, off the Internet. However, it is important to note that generation happens as a response to the prompting of the human author, David Guetta, looking to manifest a specific vision.

---

[38] See the comments of Davey Whitcraft, who stated that he "use[s] generative AI tools to experiment push forward the boundaries of human creativity." Paulius Jurcys, 'Creativity in the Age of AI' (*YouTube*, 15 May 2023) <https://www.youtube.com/watch?v=9qoY19B6MJk> accessed 30 August 2023.

[39] See comment by Sam Altman at the 2023 Bloomberg Technology Summit: "[t]here is this sci-fi idea that those [AI] systems can better address themselves, can discover new architectures, can write new code. I think we are miles away from that; but it's worth paying attention to." 'OpenAI CEO Sam Altman on the Future of AI' available at:
<https://www.bloomberg.com/news/videos/2023-06-22/openai-ceo-sam-altman-on-the-future-of-ai>
accessed 30 August 2023.

- ***Refining.*** Once he received the generated content, Guetta edited the received content and integrated it with some music. In this respect, refining is not simply the editing of a human-created text/picture (i.e., part of a process generation) but a distinctive stage of working with something given back to the creator by the third-party generative AI. In this respect, it does differ from our artist. Refining, in the sense used here, includes a broad spectrum of activities ranging from a crude copy-pasting (e.g., a lazy, dishonest student using ChatGPT to "write" their report) to a more sophisticated bundle of processes, including curation, collation, compilation, and assemblage. It also may involve re-visiting the prompting and generation stages, and *the simple sequential model introduced above may become a more dynamic, cyclical, and iterative process of ongoing prompting, generating, and filtering*. We don't know exactly, based on the Instagram post, how this worked in the Guetta case, i.e., whether he relied on the first versions produced by the AI, but the general point about the possibility and centrality of refining still stands. And this seems particularly relevant in the context of recent generative AI (ChatGPT, for instance) where the AI remembers the context from an earlier phase of the interaction. This is a recent feature and might be thought of as refining-by-design, i.e., refining is a core feature built into the operation of the technology that allows the AI to improve its performance over time *based on a process of iteration and learning*.[41] The final, approved result of refining is what we would call output. Think of it as the created piece of music in the Guetta case.

---

[40] Bruno Latour, *Reassembling the Social: An Introduction to Actor-Network Theory* (OUP 2005) 72.

- ***Deployment.*** When, where, and how the output material is used is subject to a high degree of external (i.e., human) influence. The generative AI, for the moment, cannot determine when, where, and how its creations will be used. This points to a more general feature of the current state of this technology, namely that it has no understanding of what it is creating or doing and lacks the quality of self-consciousness or individual autonomy. Again, however, the act of deployment is nothing new, and creators – including our artist – have always exerted some degree of autonomous control over whether and how their works are first published, disseminated, and used, even if that control was never total (intermediaries, such as critics and galleries in the case of painting, play a crucial role, for instance) and non-sanctioned uses frequently occur, hence the need for IP protection.

The results of the comparison are illustrated in *Figure 2*. The different actors and technologies that they utilise are added. Think of these two graphics as illustrations of the network of

---

[41] See Ari Seff, 'How ChatGPT is Trained' (*YouTube*, 25 January 2023) <https://www.youtube.com/watch?v=VPRSBzXzavo> accessed 13 February 2023.

"actors" – understood in the broad sense of human, corporate and machine actors – that constitute a creative work in any given case.

*Figure 2. Mapping Creativity in the Two Use Cases*

*Figure 2.1. Creativity in the "Painting" Case*

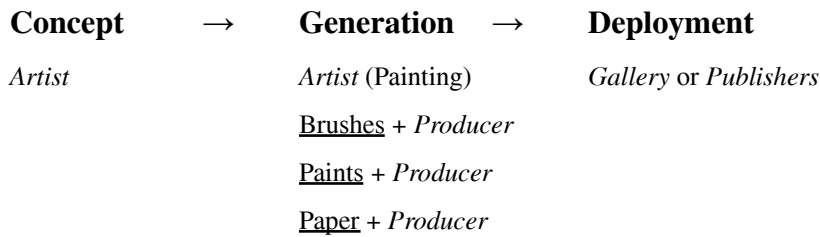

*Figure 2.2. Creativity in the David Guetta Case*

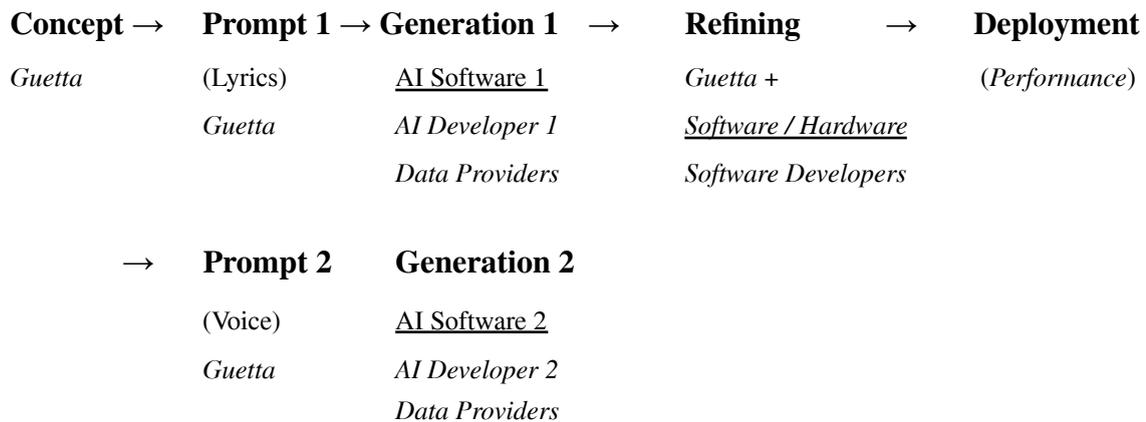

What becomes clear from the above is that the sophistication of the technologies and the network of actors (human, corporate *and* machine) is broader and more complex in the case of David Guetta. The degree of involvement of third parties in the generative moment (the AI software developer and the data they rely on) does, therefore, seem different from the painting example. We can question whether the creator has the same role in the creative process. But again, this appears to be a question of degree rather than a qualitative difference. All of this is simply to say that modern Romantic notions of creativity as purely human-generated involving creation *ex nihilo* need to be replaced by a vision of creativity as the multi-staged effects of complex networks of human, corporate, and machine (both software and hardware) "actors." The Romantic fiction of the lone genius creator is a fantasy that retroactively obscures the cultural and technological tradition implicated in all creative works. And this observation is certainly not meant to denigrate the role of creators as they play an essential role in driving a tradition forward in new and unexpected ways *through* their novel use of technology and re-imagining of tradition. Rather, our point is to suggest that the imposition of constraints upon such creative re-imagining of a tradition and use of technology should be minimised and only introduced when there is a compelling justification.

## 3. Assessing Originality in Modern US Copyright Law

Having described the model of creativity in the Guetta case, we now turn to copyright. This section briefly introduces the two requirements that form the basis of the "originality" test in modern copyright law, at least in a US context. We start with the fundamental idea/expression distinction (Section 3.1.) and then explore how the US courts have interpreted the twin elements of the "originality" test (Section 3.2.). Section 4 then analyses the originality of Guetta's composition as characterised in Section 2 above using this framework.

### *3.1. A Tangible Medium of Expression*

One of the foundational principles of modern copyright law is that ideas are not protected by copyright law and that only expressions can be copyrighted. In the US, for example, the Copyright Act provides that copyright protection does not extend "to any idea, procedure, process, system, method of operation, concept, principle, or discovery."[42] In other words, ideas and facts are not protected by copyright law; only the way in which an author expresses ideas or facts can be protected.[43] This decision that ideas receive no protection, which is now codified in Federal Law, was settled in the 1879 case, *Baker v. Selden*. The plaintiff's book describing a method of accounting was not protected, because "[w]here the truths of a science or the methods of an art are the common property of the whole world, *any author* has the right to express the one, or explain and use the other, in his own way" (our emphasis).[44]

---

[42] 17 US Code § 102(b). See also Pamela Samuelson, 'Why Copyright Law Excludes Systems and Processes from the Scope of Its Protection' (2007) 85 Tex L Rev 1921.

[43] *Nichols v Universal Pictures Corp*, 45 F2d 119 (2d Cir 1930).

There are three main reasons for this. The first is pragmatic and evidential: How can we establish *who* had *what* ideas *when*? Without any tangible evidence expressed in tangible form (e.g., paint, paper, vinyl), it is impossible to say that Person A had the idea independently from Person B.[45] This means that if there is a dispute between people accusing each other of stealing each other's ideas, it is practically impossible to determine the fact of copying without any tangible evidence, i.e., the administrative costs of resolving such disputes are prohibitively high. Furthermore, the emphasis on the function of the fixation (i.e., that the can be perceived or retrieved) rather than on form (e.g., on paper, on canvas, in writing) was to ascertain that the rights of authors are unaffected by the advancements in technology that could not have been anticipated by the legislator.[46]

A second reason that ideas do not enjoy copyright protection is the notion that limiting the use of ideas would have an inhibiting effect on scientific development and progress. This notion can be found in Article I Section 8 of the US Constitution, for example.[47]

A third reason for excluding ideas from copyright protection concerns freedom of speech. In the *Eldred* case, for example, the United States Supreme Court suggested that the idea/expression distinction is necessary to reconcile copyright law with the principle of freedom of speech embodied in the First Amendment of the Constitution.[48]

---

[44] *Baker v Selden*, 101 US 99 (1879).

[45] *Nichols v Universal Pictures Corp*, 45 F2d 119, 121 (2d Cir 1930).

[46] Dan Hunter, *Intellectual Property* (OUP 2012) 35.

[47] Art I S 8 of the US Constitution provides that the Congress has the power, "To promote the Progress of Science and useful Arts, by securing for limited Times to Authors and Inventors the exclusive Right to their respective Writings and Discoveries."

[48] *Eldred v Ashcroft*, 537 US 186, 219-220 (2003) ('[C]opyright law contains built-in First Amendment accommodations. First, it distinguishes between ideas and expression and makes only the latter eligible for copyright protection.'); *Golan v Holder,* 565 US 302, 329 (2012). Critically, see Neil W Netanel, *Copyright's Paradox* (2008) 4.

*3.2. The Originality Test*

To be protected under copyright law in the United States, a work also must be original. The originality of a work is determined by considering two factors. First, the author must create the work independently. This means that if two people create an identical work (let's say they write an identical verse or draw an identical image of a cat), each of their works can be protected under copyright even though they are completely identical. As long as the author creates their works separately, the US courts will consider it sufficient to meet the requirement of independent creation.[49]

The second element of originality is that a work must be somewhat creative. This is also known as the "modest creativity" requirement. The US Copyright Act clearly states that copyright protection "does not extend to any idea, procedure, process, system, method of operation, concept, principle, or discovery" (S. 17 USC 102(b)). In other words, naturally occurring events, real-world facts, and mathematical formulas are considered matters of the physical world. Therefore, if an author writes down something that is considered factual, such content will not be considered original.

Probably the most well-known illustration of the "modest creativity" test could be drawn from the *Feist* case decided by the US Supreme Court in 1991.[50] In this case, a dispute arose between two companies publishing telephone directories where Rural accused another company of copying Rural's white pages without permission. The Court held that a "sufficient amount of originality" is required for works eligible for copyright protection. In applying this test, the Court held that the mere arrangement of publicly known facts in alphabetical order is not creative – i.e., anyone could have done it.

So how do we explain the "modest creativity" required for works to be subject to copyright protection? The *Feist* case could be a good starting point to reverse engineer the definition of "originality." To be considered as "original," the work must contain something more than a mere collection of real-world facts. According to *Feist,* spending some time to arrange such facts in some obvious fashion (e.g., writing names of town residents alphabetically) does not meet the threshold of "modest creativity"; there must be something more that emanates from a person. Besides, whether a work is modestly creative is a question of fact in every single case, and it is for lawyers or judges to decide based on circumstances of the case at hand. As long as the work is the result of some creative endeavour by a human being, it will be deemed "modestly creative" and thus original under the US copyright law.

---

[49] "To qualify for copyright protection, a work must be original to the author," which means that the work must be "independently created by the author" and it must possess "at least some minimal degree of creativity." *Feist Publications, Inc v Rural Telephone Service Company, Inc*, 499 US 340 (1991) 345.

[50] *Feist Publications, Inc v Rural Telephone Service Company, Inc*, 499 US 340 (1991).

It should be noted – and emphasised – that the originality threshold of US copyright law is very low. Basically, according to *Feist*, anything that rises above the threshold of facts, formulas, and natural discoveries and can be attributed to an individual's creative expression can be protected by copyright. (This paper does not address the theories that might expand our understanding of why copyright exists).

Originality does not, therefore, require the author's intent to be original.[51] How come? Let's discuss a couple of examples. Imagine you are trying to bake a cake, but something in the process goes hilariously wrong. One example could be a failed Easter Bunny case that was photographed at one of the major US grocery chains Safeway and later appeared on *Cakewrecks.com* website (*Image 1*).

---

[51] See Compendium of the US Copyright Office Practices (Chapter 300), S 310.5 <https://www.copyright.gov/comp3/chap300/ch300-copyrightable-authorship.pdf> accessed 12 February 2023.

*Image 1. The Easter Bunny Cake*[52]

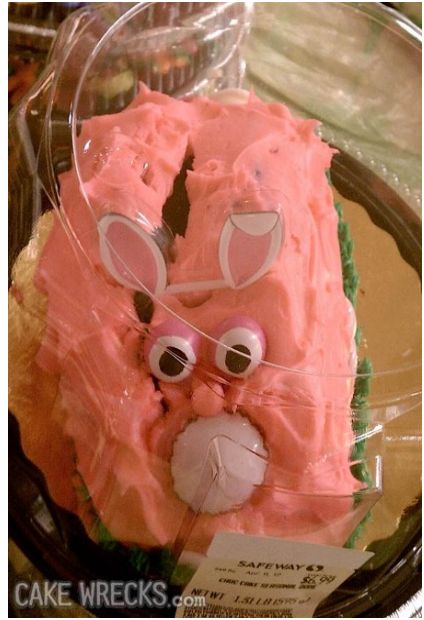

Looking at this photo, we may wonder: Did the creator intend to make such an odd-looking cake? One presumes not. What if the baker had an idea or concept of a "perfect" Easter Bunny cake but failed to materialise her vision? Would this cake still be considered original?

US courts generally agree that the author's intent to be original is irrelevant in deciding whether the work should be copyrightable. Instead, it is important to look at the outcome of the author's activity. Copyright protection is available to works that are sufficiently distinguishable regardless of whether an author tries to create something of her own. The Easter Bunny cake shown above is unique in its appearance – clearly, it is one of a kind. Also, very few people would be able to bake such a version of an Easter Bunny. Thus, it is going to be considered as original. There may be situations where a work is distinguishable from other similar works because of the author's bad eyesight, trembling muscles, or a shock caused by a clap of thunder. If an author adopts the work as her own, the copyright protection will, nevertheless, attach.[53]

---

[52] Luara Northrup, 'Safeway Doesn't Want You Photographing Their Terrible Cakes For That Blog About Terrible Cakes' (*Consumerist*, 13 November 2012).
<https://consumerist.com/2012/11/13/safeway-doesnt-want-you-photographing-their-terrible-cakes-for-that-blog-about-terrible-cakes/> accessed 12 February 2023.
[53] Opinion by Judge Jerome Frank in *Alfred Bell & Co v Catalda Fine Arts, Inc*, 191 F2d 99 (2d Cir 1951).

How about those works that do not appear appealing to the eye? Should copyright protect works we may find not pretty, ugly, or even insulting? In the US and many other countries, copyright practice adopts a principle of aesthetic neutrality which means that lawyers (i.e., attorneys, judges, or government agencies) should not stand in judgment as to whether works are beautiful and aesthetically pleasing or not. Beauty is in the eye of the beholder, after all. The Easter Bunny cake will be deemed to meet the threshold of minimum creativity, even if most people find it distasteful or otherwise unattractive.

Finally, originality does not mean that the work must be new. You can obtain copyright protection even if the work – let's say a phrase – is not new. If it is "modestly creative" (and we now know this threshold is very low), then you can obtain the copyright in the work. The crucial point here is the distinction of the terminology used when artistic works are protected by copyright and technological inventions protected under patent law. An invention can be patented only if the three requirements of novelty, inventive step and practical utility are satisfied. To determine whether an invention is new, patent examiners usually check all available databases and determine whether the invention is new: has anyone already come up with this idea? Has this type of invention been patented already? If patent examiners determine that the idea in the patent application is "new", then the invention can be protected by a patent provided that two other requirements are satisfied.

As such, novelty is not required to benefit from copyright protection: the work must simply be original. If the author creates something independently, and it is not a literal copy of another work, such work can be granted protection under copyright law.

## 4. The Originality of Works Created with Generative AI Tools

Much ink has been spilled discussing how AI will change everything. In the creative and legal domains, there are many discussions about whether "AI-generated" works should be protected by copyright.[54] Here, we want to make a case that the existing framework, as laid out in Section 3 is sufficiently durable to handle the mode of creativity described in Section 2. Considering what we already know about assessing the originality of works, can the composition of David Guetta be deemed original? Was this work created independently? Does the composition meet the "minimum creativity" threshold? Should works be considered original, and thus subject to copyright protection, when created using generative AI tools such as Midjourney or ChatGPT? Let us address the elements of originality in turn.

### *4.1. Independent Creation*

The question of whether a work is independently created most frequently arises in copyright disputes where someone is accused of copying a plaintiff's work.[55] It would then be up to the defendant to prove that they created the work independently, without having seen or heard the plaintiff's work. If we simplify it, their argument would be, "I created it on my own." So, in David Guetta's words, he would argue that he was sitting alone in his studio and came up with the concept to try and see what happens when he mixes Eminem's rap with the AI-generated music which we would get by adding the prompt "the future of rave." It is David Guetta and no one else who came up with the prompts for the text and music.

---

[54] See eg, Yang Xiao, 'Decoding Authorship, Is There Really no Place for an Algorithmic Author Under Copyright Law?' (2023) 54 IIC 5; Michael D Murray, 'Generative and AI Authored Artworks and Copyright Law' (2023) 43(1) Hastings Comm & Ent L J 28; Begona Gonzalez Otero, 'Machine Learning Models Under the Copyright Microscope: Is EU Copyright Fit for Purpose?' (2021) GRUR Int 1043; Yong Wang and Hongxuyang Lu, 'Copyright protection for AI-generated outputs: The experience from China' (2021) 42 CLSR; Jane Ginsburg, 'People Not Machines: Authorship and What It Means in the Berne Convention' (2018) IIC 131; Giancarlo Frosio, 'The Artificial Creatives: The Rise of Combinatorial Creativity from Dall-E to GPT-3' <https://papers.ssrn.com/sol3/papers.cfm?abstract_id=4350802> accessed 29 August 2023; Ryan Abbott and Elizabeth Rothman, 'Disrupting Creativity: Copyright Law in the Age of Generative Artificial Intelligence' <https://papers.ssrn.com/sol3/papers.cfm?abstract_id=4185327> accessed 29 August 2023; Jon McCormack et al, 'Is Writing Prompts Really Making Art?' <https://arxiv.org/abs/2301.13049> accessed 23 February 2023; Jane C Ginsburg & Luke A Budiardjo, 'Authors and Machines' (2019) 34 Berkeley Tech L J 343; Vicenç Feliú, 'Our Brains Beguil'd: Copyright Protection for AI Created Works' (2021) USF Intell Prop & Tech L J 105; Carys Craig and Ian Kerr, 'The Death of the AI Author' (2021) OLR 35; Daniel J Gervais, 'The Protection Under International Copyright Law of Works Created with or by Computers' (1991) IIC 628; Pamela Samuelson, 'The Future of Software Protection: Allocating Ownership Rights in Computer-Generated Works' (1986) 47 U Pitt L Rev 1185.

[55] This issue of independent creation is conceptualised as an affirmative defence, see *Feist*, 499 US 340 (1991) 345-46.

One of the essential points of independent creation is that the work emanates from a human being and that the work is created by a natural person.[56] However, and as emphasised above, people always use some kinds of tools to achieve their desired results in the real world. Leonardo da Vinci used a palette to paint the Mona Lisa. The baker of the Easter Bunny used various ingredients, presumably a spoon, a spatula, and an oven to (albeit ineffectually) bake the cake. Photographers and graphic designers use Adobe Photoshop to edit their images. Typically, David Guetta would use a computer, DAW software, software plugins, turntables, and a collection of vinyl recordings to create music. But this time he also used generative AI tools (ChatGPT and Uberduck) to create his remix. That all seems totally normal and unproblematic.

When we talk about generative AI tools in their current phase of adoption, it should be noted that prompts may need to be modified, clarified, and adjusted many times to get usable and satisfying results. In a generative AI environment, the content does not emerge from nowhere but must be initiated by a person who types in a prompt or prompts, and then possibly spends time improving the results by further modifying the prompts as part of the process of refining.[57] And just like in the baking of the Easter Bunny cake example, it is up to each individual to choose which combination of tools and ingredients to use and choose the colour of the cake, so to speak. David Guetta would appear to have a strong case that the prompts he used contributed to an "independent creation" and that the "independent creation" element of the originality test is, therefore, met.

More generally, the question then arises of how the creative process differs between using conventional tools (e.g., a brush) and generative AI tools. Intuitively, we might assume that the creator has less control over the final output when machine learning and data are used to generate content. But it seems that the question here is more about the *degree* of human involvement (as opposed to the input from the technology) in the creative process rather than any qualitative difference. Again, a Romantic conception of creative *ex nihilo* may affect (distort) our judgment on this point, as it raises the threshold beyond the traditional legal standard.

---

[56] *Alfred Bell & Co v Catalda Fine Arts, Inc*, 191 F2d 99, 102 (2d Cir 1951) ("'Original' in reference to a copyrighted work means that the particular work 'owes its origin' to the 'author.'")

[57] See comments by Davey Whitcraft who noted that creating truly amazing works with generative AI tools requires much time (weeks, or months), Paulius Jurcys, 'Creativity in the Age of AI' (*YouTube*, 15 May 2023) <https://www.youtube.com/watch?v=9qoY19B6MJk> acccessed 30 August 2023.

As observed in Section 2, it is important to avoid the over-simplistic suggestion that with the emergence of generative AI, we are moving *from* a world of human-driven creation *to* a world of machine-driven creation. Instead, our suggestion is that creation has always been the product of human-machine hybrid forms of collaborative creation – what we might characterise as a *co-creation* of a human author or authors (in a case where there are multiple creators) *and* technology – and that generative AI is merely the latest and most sophisticated iteration of such a trend. Moreover, as highlighted in Section 2, multiple third parties are implicated in any creative act in a contemporary context where a wide and diverse range of companies produce the technologies that are utilised by creators. The number and degree of involvement of third-party actors involved is much greater in the Guetta case, for example.

This point is particularly pertinent in the context of generative AI. Machine learning and generative AI software of the kind used by David Guetta relies on enormous quantities of data accumulated, typically, off the Internet and which the third-party AI developer oftentimes does not have any clear or obvious right to use. Certainly, they do not have the permission of the original creators or owners of such data. Nevertheless, generative AI completely depends on this dataset to produce anything of value.[58] However, as we further elaborate in Section 5.2., the determination of independent creation should not depend on the question of whether the third-party developer of generative AI tools has used the training data lawfully.

---

[58] For a discussion, see Jenny Quang, 'Does Training AI Violate Copyright Law?' (2021) 36 Berkeley Tech L J 1401. Watermarking has been explored as a possible solution to prevent copyright violations by ML models, see eg, Sofiane Lounici, 'Protect your machine learning models with watermarking' (*SAP*, 8 November 2021) <https://blogs.sap.com/2021/11/08/protect-your-machine-learning-models-with-watermarking/> accessed 23 February 2023 and 'Protecting the Intellectual Property of AI with Watermarking' (*IBM*, 20 July 2018) <https://www.ibm.com/blogs/research/2018/07/ai-watermarking/> accessed 23 February 2023.

*4.2. Modest Creativity*

The element of "modest creativity" in determining the originality of a work raises more intricate questions. How can we apply "modest creativity" in a generative AI context? At which point in the creative process should we assess whether the "modest creativity" test is met? Should the prompt itself be original, or should we look at the final version of the work created using "generative AI" tools (what we characterised as the output in Section 2)? It is not our intention to suggest that these can be easily answered in all cases – and we would accept the Guetta case is a relatively simple one. Moreover, harder cases may involve difficult judgments about whether to offer protection at all or the form of protection and its amount. Nevertheless, courts are routinely confronted with difficult and delicate questions of this kind, and we would cautiously suggest that expert courts in concrete cases are well or, at least, best placed to answer such questions.[59] And even if the courts get these decisions wrong from time to time, such a system seems preferable to the imposition of a crude rule that limits the use of technologies and stifles opportunities for creativity.

In *Feist*, for example, the US Supreme Court held that to benefit from copyright protection, works have to rise above being mere facts of the physical world. But as we discussed before, the threshold of "modest creativity" is very low, at least under the US copyright law. Furthermore, in assessing what is "modestly creative," US courts only analyse the outcome of the creative process, or the actual expression, that a person came up with (not the author's intention).

What do these principles covered by the "modest creativity" element suggest in our analysis of the clip created by David Guetta with generative AI tools? If the author's intentions do not matter, it probably means that we do not need to analyse whether Guetta's prompts were original or creative; rather, we would probably need to focus on the question of whether the final work meets this threshold of "minimum creativity." We would posit that the "Future Rave with Eminem" piece created by Guetta clearly rises above a mere compilation of facts (the telephone directory of *Feist*). Regardless of what Guetta had in mind or what his intentions were when he started out, the output clearly and obviously meets the threshold of being "somewhat creative."

---

[59] Richard Lawler, 'The US Copyright Office says you can't copyright Midjourney AI-generated images' (*The Verge*, 23 February 2023) <https://www.theverge.com/2023/2/22/23611278/midjourney-ai-copyright-office-kristina-kashtanova> accessed 23 February 2023. For a critical analysis see also Van Lindberg, 'A mixed decision from the US Copyright Office' (*Process Mechanics*, 22 February 2023).

What about the fact that Guetta did not know what the result would be at the concept stage of the creative process? Does this matter in assessing originality? As highlighted with our Easter Bunny Cake example, unexpectedly created works – be they successful and appealing to the eye, or a disastrous blunder – can still be protected by copyright. Following this line of reasoning, the fact that we *cannot* fully predict *ex ante* what results the generative AI *might* deliver does not matter in making a final assessment as to whether the work is "modestly creative" or not.

One might also argue that such works are modestly creative because they are created using tools that have not existed before. The second element of the originality test means that the work merely possesses *some* minimum level of creativity. What David Guetta has created using generative AI tools is a piece of music that did not exist before. He devised an *original* concept to combine Eminem's voice with FR rave music and reimagined what the future rave might look like. Yes, indeed, he used AI tools to materialise his vision, but it was his idea to create this kind of mix. He also used those AI tools himself and refined the results to create a unique output that he deployed in his show. A new concept. An iterative process of creation. New content, new beats, and the crowd loved it!

## 5. Copyright Re-Visited

When it comes to creative works, isn't it always the case that artists try to create something that gains the attention of a mass audience and generates media interest? Writers dream of landing publishing contracts with major publishers and making it to the New York Times bestseller list. Visual artists aspire to create the next Mona Lisa. It is natural that machine learning and generative AI tools make everyone excited because they expand and share this feeling of being capable of creating something unique and beautiful. And, from that perspective, don't all artists always aspire to create something truly original?

One of the key advantages of generative AI is that it accelerates the process of creativity: rather than spending time in front of an empty sheet of paper, we can now prompt AI to provide us with some initial ideas and possible suggestions that provide impetus to the process of creating.[60] From that perspective, generative AI tools are appealing to people working in creative industries. For individual users businesses, generative helps save time and curtail the costs of production.[61] People who are worried that AI tools will replace humans should, perhaps, worry less and learn how to use those tools to increase their own productivity.[62] Even if we don't completely rely on such tools, they stimulate creativity and – when combined with a process of refining – they are opening new opportunities and possibilities.

## *5.1. Identifying the Human Author-in-the-Loop*

Against this emerging background, what we might label "AI literacy" becomes vital. There is no turning back, and our society must find ways to cultivate the ability to work with these technologies in a responsible and effective manner. In order to better understand how to adjust the existing legal frameworks to the world where generative AI technologies are ubiquitous, it is important to understand in what ways those new AI tools change the well-known creative process that copyright law aims to incentivize and reward. In light of what we discussed about the creative process where human authors use generative AI tools in Section 2 above, we believe that in most cases, it is not very difficult to identify at which stages of the creative process a human author is involved.

---

[60] Reid Hoffman, *Impromptu: Amplifying Our Humanity Through AI* (Dallepedia 2023) 18.

[61] Ibid 110; Mark A Lemley, 'How Generative AI Turns Copyright Upside Down' (2023) <https://papers.ssrn.com/sol3/papers.cfm?abstract_id=4517702>, accessed 29 August 2023 who discusses the concept of "cheap creativity."

[62] See Hoffman, supra n 60, 49-50, who makes a comparison between creating in the style of John Lennon, and asking how would John Lennon use this tool.

We also see how the unpacked notion of the creative process, which we described in Section 2, provides a clear response to those who have been questioning whether the notions of independent creation and modest creativity are of any use in the age of generative AI. In the copyright law literature, the idea of doing away with the requirement of independent creation is not entirely new,[63] and it seems to emerge again in the context of generative AI. As explained previously, one of the key functions of the "independent creation" requirement is to identify that there was a human being who came up with an idea and created something that is fixed in tangible form. Hence, we are of the opinion that the notion "AI-generated" work is inaccurate and leads to confusion; it is misleading because, as we showed in Section 2, there is always a human/author in the loop, also in those cases when creators utilize generative AI tools.

Curiously, the US Copyright Office has adopted a more cautious approach on this point requiring artists to disclose which parts of a work have been created using generative AI tools.[64] Pursuant to this "prompt-based approach", the US Copyright Office ignores the creativity contributed by the AI system and focuses on rewarding creativity contributed by human creators, provided that human authors are able to show how the series of prompts rise to meet the threshold of minimum creativity.[65]

---

[63] Christopher Buccafusco, 'There's No Such Thing as Independent Creation, and it's a Good Thing, Too' (2023) 64 Wm & Mary L Rev 1617.

[64] See Copyright Office Statement of Policy, 88 Fed Reg 16192 (16 March 2023) <https://www.govinfo.gov/content/pkg/FR-2023-03-16/pdf/2023-05321.pdf> accessed 29 August 2023: "In the case of works containing AI-generated material, the Office will consider whether the AI contributions are the result of 'mechanical reproduction' or instead of an author's 'own original mental conception, to which [the author] gave visible form.' The answer will depend on the circumstances, particularly how the AI tool operates and how it was used to create the final work."

[65] There is at least one lower court decision in the District court of Washington DC which seems to have followed this approach. See *Stephen Thaler v Shira Perlmutter,* Case 1:22-cv-01564-BAH; Riddhi Sethi & Isaiah Poritz, 'AI-Generated Art Lacks Copyright Protection, D.C. Court Says (1)' (*Bloomberg Law*, 18 August 2023) <https://news.bloomberglaw.com/ip-law/ai-generated-art-lacks-copyright-protection-d-c-court-rules> acccessed 30 August 2023; Heather Whitney, 'Court Says No Human Author, No Copyright (but Human Authorship of GenAI Outputs Remains Uncertain)' (Technology & Marketing Law Blog, 22 August 2030) <https://rb.gy/7hmaz> acccessed 30 August 2023.

Although such an approach could be considered as a practical step forward, it could be criticised from at least three possible angles. First, until now, artists have never been required to disclose the tools used in the creative process in copyright application forms in the US, but a view seems to be gaining traction that now might be the time to reverse this and adopt new requirements. One can argue whether generative AI is all that different from other digital creation tools such as Adobe Photoshop to warrant such a historic introduction of disclosure requirements.[66] Second, it may be questioned whether such disclosure requirements are in line with the no formalities principle adopted in Article 5(2) Berne Convention.[67] Finally, not only contemporary copyright law is not well designed to support a prompt-based system,[68] it also seems like a dangerous trend that misconceives the character of the creative process in the case of generative AI.

In mass media publications, it is easy to catch the attention of wide audiences with such and bold statements that certain legal institutions do not meet the test of time and that technology moves faster than the law. But this has always been the case, and laws and social norms are constantly adjusted to meet the test of rapidly evolving technologies.

With regard to the issue of originality in copyright, we would propose a more modest approach against the quick and easy dismissal of legal concepts that have evolved and stood the test of time. Looking back at history, technology has always preceded the existing copyright laws (think of the printing machine, the invention of cars, VCRs, live audio, and video streaming).[69] And, as we hopefully demonstrated here, the two elements of originality developed by courts in solving real cases prove to be helpful in approaching David Guetta's music composition. We would modestly suggest that they might also be helpful in other cases where similar questions arise. We would concede that drawing a bright line is hard and that – as mentioned already – it will require difficult and delicate decisions. Nevertheless, this is preferable to the alternative – denying protection completely – which risks throwing the baby out with the bath water and placing the artists at the cutting edge of creativity in a weak position.

---

[66] See comments by Katrina M Kershner, Paulius Jurcys, 'Creativity in the Age of AI' (*YouTube*, 15 May 2023) <https://www.youtube.com/watch?v=9qoY19B6MJk> accessed 30 August 2023.

[67] Article 5(2) of the Berne Convention states, "[t]he enjoyment and the exercise of [the copyright owner's] rights shall not be subject to any formality." For a broader discussion, see Jane C Ginsburg, 'Berne Forbidden Formalities and Mass Digitisation' (2016) 96 B U L Rev 745; David R. Carducci, 'Copyright Registration: Why the U.S. Should Berne the Registration Requirement ' (2020) 33 Ga St U L 873.

[68] Lemley, supra n 61.

[69] See e.g., Mark Rose, *Authors in Court: Scenes from the Theater of Copyright* (Harvard Univesity Press, 2016). For an overview of how technologies and underlying business models have affected the music industry, see Allen Bagfrede & Cecily Mak, *Music Law in the Digital Age* (Berklee Press 2009).

And finally, the David Guetta story reveals something more general about creativity in the digital age. Earlier – and influential – models of creativity as purely human-driven and involving creation *ex nihilo* do become harder to sustain in a new age of generative AI. The creator – understood as an abstract or pure will operating independently of any technology – has always been a fantasy and never existed. Creation has always been over-determined by a collaborative combination of humans *and* machines. We need to be more attuned both to hybridity and the role of the nonhuman in the constitution of creativity. Hybrid-networked (i.e., human – corporate – machine) *creators* have always created hybrid – networked cultural *forms* (i.e., creations that blend human and technology-constituted elements). But – we would suggest – this hybridity becomes increasingly visible and complex in the context of a world of generative AI. Crucially, however, as we have suggested here, this does not mean copyright or notions of originality are redundant or copyright is unable to accommodate Guetta and other similar cases. Quite the contrary, notions of originality provide a robust framework for managing new technologies and the new forms of creativity that such technologies facilitate.

## 5.2. Does the Lawfulness of Training Data Affect Originality?

An important complicating factor in the case of generative AI is the direct dependence of the AI systems on aggregated data used to train the AI models. One question that needs to be addressed is whether the use of publicly available data for training purposes is justifiable or not. More precisely, we need to ask whether the reliance on such data by generative AI is sufficiently different from earlier technologies to justify excluding the works that are created using generative AI tools from the scope of copyright protection.

Our intention is not to answer this question here but merely to provide greater clarity and precision to the issues that need to be resolved as we move forward. It is important that this question of training data does not "crowd out" the discussion about the key justifications of copyright law and originality; questions of copyright should not be reduced to this issue of the data used to train underlying models that power AI solutions.

This danger seems particularly unfortunate when combined with an account that locates the key moment of creativity at the point where the black box of the AI systems generate content. A privileging of this aspect of the creative process, seems likely to result in a myopic emphasis on the status of the training data, whereas a more nuanced account of creativity shifts the focal point of the discussion towards the issue of originality and human authorship explored here. And just to reiterate, this is not to deny the importance of these questions, merely to highlight how the black box moment is only one – albeit very important stage – in a more nuanced and complex process (*Figure 3*).

*Figure 3. Situating Creativity*

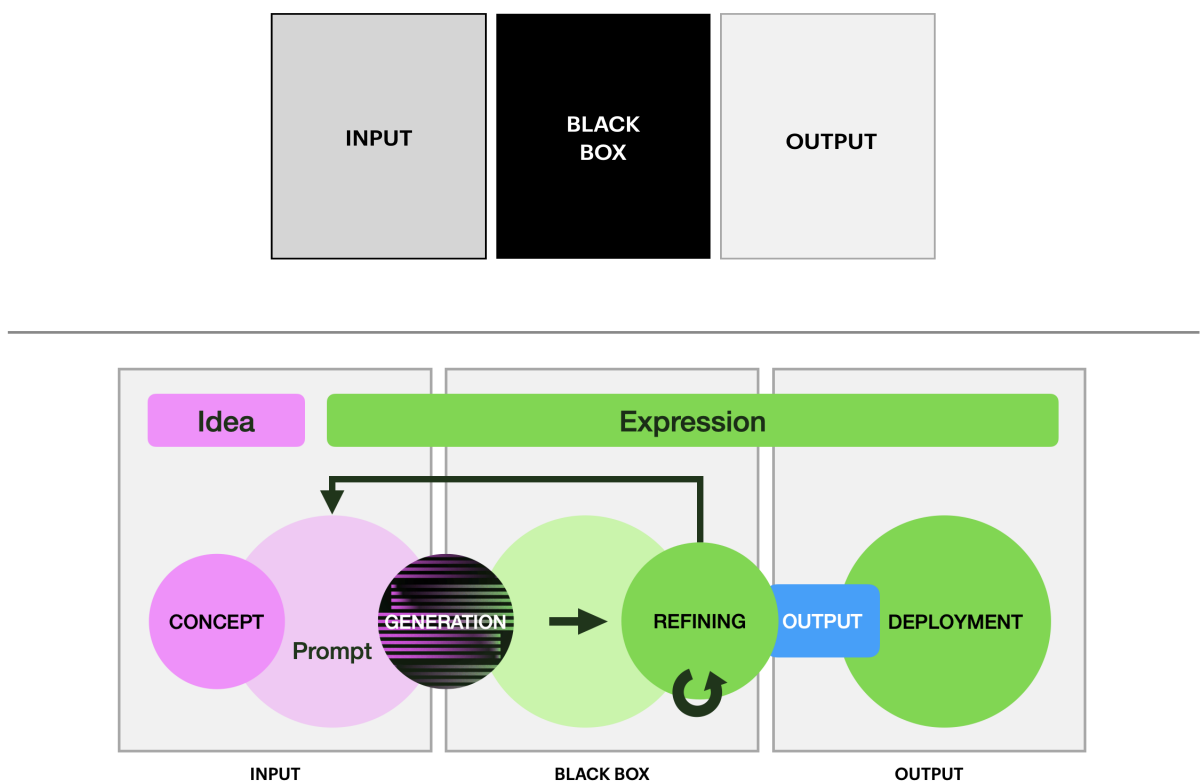

Another distinct question that we would like to highlight here relates to the question of whether the legality of the use of publicly available data for training purposes should be included in assessing the originality of the output created by the user of the generative AI platform. Moreover, and crucially, when we talk about consumer-facing AI platforms like Chat-GPT or Midjourney, it seems unfair that the question of the lawfulness of the use of training data, in general, is added to the copyrightability assessment of an individual work created by an individual *using* that generative AI system. More specifically, the burden (and cost) for sorting out the legality of the training data conundrum should be placed on the shoulders of the platform providing the generative AI system rather than the individual user that merely used it as one element of the creative process as described in Section 2 above.

A final consideration as to the relevance of the training data is whether the generative AI tools are used by an individual user/creator or whether generative AI models are used in a corporate environment by employees in enterprises. It is easy to envisage several different scenarios involving business, which require further discussion. For example, where enterprises train their own models based on the company's own data (for example, an architecture or interior design firm training its own image generation model). In such a case, the costs should be internalized by the firm. Another case, when one firm uses another firm's generative AI platform, a contractual arrangement between the parties might be necessary, with companies relying on insurance to protect against any potential legal or business risks.

## *5.3. Policy Choices in Fringe Cases*

Another possible example where the issue of so-called "AI authorship" is raised relates to content generated from snippets of videos captured by cameras embedded in various IoT devices (e.g., Ring doorbell cameras or smart pet feeders such as Furbo). Such smart IoT devices are initially installed by their owners who buy and place these devices in their preferred locations: smart doorbells with cameras are usually installed at the entrance door, while pet feeders are usually positioned in the location of the room where the camera is able to capture the "best view" of the space. Such smart IoT devices also have an accompanying application, which users download onto their mobile phones and can "opt-in" to get regular notifications and stitched videos capturing the highlights of the day, a week, or a month.

The legal question is whether such content is copyrightable or not? And, if so, who owns the content? The device maker? The device owner? Nobody?[70]

To address these questions, we would like to make two observations. First, the videos that are captured, stored, and processed by companies selling such "smart" IoT devices should not be considered as content that is "autonomously generated by AI". While it is true, that such videos are captured by video cameras that automatically turn on when movement is detected, such snippets or compilations thereof are a result of sensors and software that are programmed to perform a specific function, i.e., to recognise, categorize, and stitch the snippets together. Although some might potentially turn out to be relevant, interesting, funny, or newsworthy, such uses of smart IoT devices equipped with motion recognition sensors and night-vision cameras are not really at the core of copyright.

Second, it is a matter of policy choice whether such content is copyrightable or should be protected under certain contractual arrangements between the device manufacturer and the owner/user of the device. Notably, there is a software layer that combines computer vision and data labeling technologies to recognize, categorize, and stitch the snippets of content together. From an institutional perspective, the responsibility to decide on copyright ownership may lie with either (a) the policymaker, who has to make a decision on this issue, or (b) the camera maker, who sets forth the solution in the terms of use, which the buyer of the device must agree to and live with. Such a policy choice with regard to copyrightability and the allocation of initial title to the content captured by smart IoT devices could be deemed as a layer of a higher-level normative framework. The debate about content captured by sensor-equipped cameras, such as capturing cats, dogs, and mailmen around houses, distracts us from the broader conversation about the purpose of copyright, which traditionally centers around human creators.

---

[70] By way of illustration, the Ring Terms of Service provide that "You are solely responsible for all of your Content (including Content you share through the Ring Neighbors feature or application). "Content" means all audio, video, images, text, or other types of content captured by Products or provided to us (including content posted by you) in connection with the Services. You represent and warrant that:
(a) you own the intellectual property rights in Content posted by you or otherwise have the right to post the Content and grant the license set forth below, and (b) the posting and use of your Content on or through the Services does not violate the privacy rights, publicity rights, copyrights, contract rights, intellectual property rights or any other rights of any person." <https://ring.com/terms> accessed 29 August 2023.
The Terms and Conditions of Furmo, a camera-equipped smart pet feeder are less clear: "Our Service may allow you to publicly share and otherwise make available certain information, graphics, videos, or other material ("Content"), including, without limitation, Content captured through or in connection with your use of the Services. … You are solely responsible for the Content that you publicly upload, transmit, share or otherwise disseminate on and through the Services, including its legality, reliability, and appropriateness." <https://furbo.com/us/pages/terms-conditions> accessed 30 August 2023.

More specifically, *The Firefighter* case in the US,[71] where a gas maintenance station employee captured a photo of a fireman carrying a baby out of a destroyed building following an explosion in Oklahoma City provides an illustration of the policy choice at stake. In this case, the question was whether the photo was a work made for hire, or whether it was a sole-authorship work? Situations might usefully be framed as law and economics issues: Who is in a better position to utilize or monetize the photo, the employee or the employer? As such, they might be best thought of as case-by-case decisions, and it is impossible for the regulator to anticipate all such scenarios. On a more general note, we again reiterate the view that videos captured by security cameras are not "works that are autonomously generated by AI." Shifting attention to that aspect to answer the question of whether it is copyrightable seems, once again, overly reductive and simplistic.

## 6. Conclusions

The main claim of this paper is that current copyright law in the US provides a durable framework that can and should manage a Guetta-like case. However, such an accommodation does raise several novel issues and questions, first, about the relationship between humans and machines in the creative process and, secondly, about the shifting character of the network of relevant stakeholders implicated in this process.

Our main argument here seeks to clarify the notion of an AI-generated work – as used in many provocative recent headlines. Our point is to emphasize that there is no such thing as purely AI-generated work, at least not yet, and that there is always a human/author in the loop of the creative. The idea of "AI-generated work" is overly simplistic and potentially misleading and does a disservice to those seeking to experiment creatively with generative AI systems. Even more so, the notion of "AI-generated work" implies that AI can act completely autonomously in generating AI content - something that is not factually true. As we try to show in the paper using the Guetta example, human/author involvement usually occurs either at the input (prompting) or output stage; and the creation of somewhat creative works usually requires human intervention and refinement of interim iterations of a work.

---

[71] *Oklahoma Natural Gas Co v Larue*, No. 97-6087 (10th Cir 1998).

As such, it is important that new concepts and terminology are developed that more accurately characterize the forms of creativity emerging in our rapidly evolving digital culture. We prefer the idea of "co-creation of works with AI tools" as this type of expression seems to capture the hybrid character of the creative process that is happening in the generative AI space right now. We believe that our effort to unfold the five dimensions of human authorship in creating works by using generative AI tools (i.e., conception, prompting, generation, refinement and deployment) can shed new light on the legal concept of *human* authorship which during the 19th and 20th centuries increasingly shifted focus from the author to the work.[72]

In this context, we might borrow Lawrence Lessig's term "remix" and think about how it might be deployed in an age of generative AI. Crucially, this concept points to an essential feature of digital culture, namely the centrality of quotation and iteration, and our struggles with imagining the future and developing new artistic forms.[73] Perhaps, the only or, at least, best response to a culture where we struggle to imagine the future is remixing and recycling – a self-conscious, often ironic, revisiting and repackaging of the past – and in that respect AI tools offer a contemporary and timely means to navigate the impasse of the current historical moment.

From a legal perspective, our purpose here has been to suggest that the low threshold of creativity in US copyright law is easily met in a case such as David Guetta's. Nevertheless, further research is needed to analyse more specific use cases involving music, voice, multiple generative art segments, as well as the use of specific applications across these different media. As such, we need to have an open and enlightened discussion about hard or fringe cases, where there may well be far less human creativity than discussed here, and a more restrictive approach may be justified.

---

[72] Peter Jaszi, 'Toward a Theory of Copyright: The Metamorphoses of 'Authorship'' (1991) Duke L J 455.
[73] Lawrence Lessig, *Remix: Making Art and Commerce Thrive in the Hybrid Economy* (Penguin 2008).

Our conclusion is that when it comes to genuine creators, who are using AI systems as innovative and unique tools to push the boundaries of their creativity, copyright law remains one of the primary mechanisms to facilitate them in this pursuit. Exaggerated claims about the role and capacity of AI in creation and possible flaws in the training data of such models should not be used as a mechanism or justification to curtail the discussion, stifle creativity, and limit opportunities to use AI tools in pushing digital culture forwards and in new interesting directions.